\documentclass[twoside,twocolumn,9pt]{article}
\usepackage{extsizes}
\usepackage[super,sort&compress,comma]{natbib} 
\usepackage[version=3]{mhchem}
\usepackage[left=1.5cm, right=1.5cm, top=1.785cm, bottom=2.0cm]{geometry}
\usepackage{balance}
\usepackage{mathptmx}
\usepackage{sectsty}
\usepackage{graphicx} 
\usepackage{lastpage}
\usepackage[format=plain,justification=justified,singlelinecheck=false,font={stretch=1.125,small,sf},labelfont=bf,labelsep=space]{caption}
\usepackage{float}
\usepackage{fancyhdr}
\usepackage{fnpos}
\usepackage[english]{babel}
\addto{\captionsenglish}{%
  \renewcommand{\refname}{Notes and references}
}
\usepackage{array}
\usepackage{droidsans}
\usepackage{charter}
\usepackage[T1]{fontenc}
\usepackage[usenames,dvipsnames]{xcolor}
\usepackage{setspace}
\usepackage[compact]{titlesec}
\usepackage{hyperref, multirow}

\usepackage{epstopdf}

\definecolor{cream}{RGB}{222,217,201}

\makeatletter
\newsavebox{\@brx}
\newcommand{\llangle}[1][]{\savebox{\@brx}{\(\m@th{#1\langle}\)}%
  \mathopen{\copy\@brx\mkern2mu\kern-0.9\wd\@brx\usebox{\@brx}}}
\newcommand{\rrangle}[1][]{\savebox{\@brx}{\(\m@th{#1\rangle}\)}%
  \mathclose{\copy\@brx\mkern2mu\kern-0.9\wd\@brx\usebox{\@brx}}}
\makeatother

\begin{document}

\pagestyle{fancy}
\thispagestyle{plain}
\fancypagestyle{plain}{
\renewcommand{\headrulewidth}{0pt}
}

\makeFNbottom
\makeatletter
\renewcommand\LARGE{\@setfontsize\LARGE{15pt}{17}}
\renewcommand\Large{\@setfontsize\Large{12pt}{14}}
\renewcommand\large{\@setfontsize\large{10pt}{12}}
\renewcommand\footnotesize{\@setfontsize\footnotesize{7pt}{10}}
\makeatother

\renewcommand{\thefootnote}{\fnsymbol{footnote}}
\renewcommand\footnoterule{\vspace*{1pt}%
\color{cream}\hrule width 3.5in height 0.4pt \color{black}\vspace*{5pt}} 
\setcounter{secnumdepth}{5}

\makeatletter 
\renewcommand\@biblabel[1]{#1}            
\renewcommand\@makefntext[1]%
{\noindent\makebox[0pt][r]{\@thefnmark\,}#1}
\makeatother 
\renewcommand{\figurename}{\small{Fig.}~}
\sectionfont{\sffamily\Large}
\subsectionfont{\normalsize}
\subsubsectionfont{\bf}
\setstretch{1.125} 
\setlength{\skip\footins}{0.8cm}
\setlength{\footnotesep}{0.25cm}
\setlength{\jot}{10pt}
\titlespacing*{\section}{0pt}{4pt}{4pt}
\titlespacing*{\subsection}{0pt}{15pt}{1pt}

\fancyfoot{}
\fancyfoot[LO,RE]{\vspace{-7.1pt}\includegraphics[height=9pt]{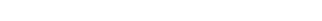}}
\fancyfoot[CO]{\vspace{-7.1pt}\hspace{13.2cm}\includegraphics{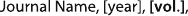}}
\fancyfoot[CE]{\vspace{-7.2pt}\hspace{-14.2cm}\includegraphics{head_foot/RF}}
\fancyfoot[RO]{\footnotesize{\sffamily{1--\pageref{LastPage} ~\textbar  \hspace{2pt}\thepage}}}
\fancyfoot[LE]{\footnotesize{\sffamily{\thepage~\textbar\hspace{3.45cm} 1--\pageref{LastPage}}}}
\fancyhead{}
\renewcommand{\headrulewidth}{0pt} 
\renewcommand{\footrulewidth}{0pt}
\setlength{\arrayrulewidth}{1pt}
\setlength{\columnsep}{6.5mm}
\setlength\bibsep{1pt}

\makeatletter 
\newlength{\figrulesep} 
\setlength{\figrulesep}{0.5\textfloatsep} 

\newcommand{\topfigrule}{\vspace*{-1pt}%
\noindent{\color{cream}\rule[-\figrulesep]{\columnwidth}{1.5pt}} }

\newcommand{\botfigrule}{\vspace*{-2pt}%
\noindent{\color{cream}\rule[\figrulesep]{\columnwidth}{1.5pt}} }

\newcommand{\dblfigrule}{\vspace*{-1pt}%
\noindent{\color{cream}\rule[-\figrulesep]{\textwidth}{1.5pt}} }

\makeatother

\twocolumn[
  \begin{@twocolumnfalse}
{\includegraphics[height=30pt]{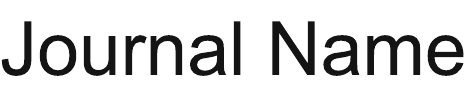}\hfill\raisebox{0pt}[0pt][0pt]{\includegraphics[height=55pt]{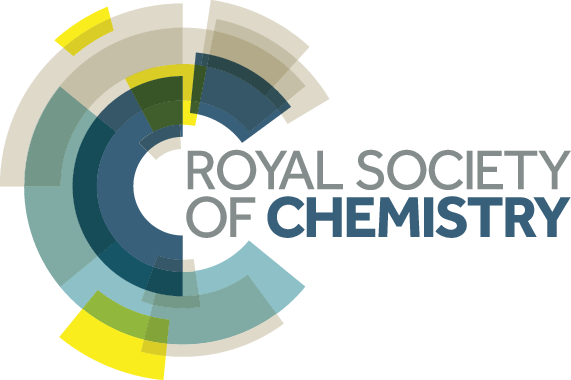}}\\[1ex]
\includegraphics[width=18.5cm]{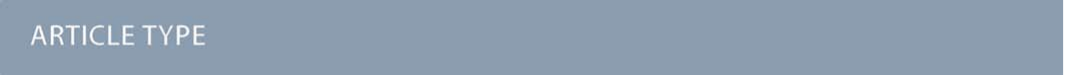}}\par
\vspace{1em}
\sffamily
\begin{tabular}{m{4.5cm} p{13.5cm} }

\includegraphics{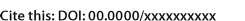} & \noindent\LARGE{\textbf{Polymer Composites Informatics for Flammability, Thermal, Mechanical and Electrical Property Predictions$^\dag$}} \\
\vspace{0.3cm} & \vspace{0.3cm} \\

 & \noindent\large{Huan Tran,$^{\ast}$\textit{$^{a}$} Chiho Kim,\textit{$^{a}$} Rishi Gurnani,\textit{$^{a}$} Oliver Hvidsten,\textit{$^{a}$} Justin DeSimpliciis,\textit{$^{a}$} Rampi Ramprasad,\textit{$^{a}$} Karim Gadelrab\textit{$^{b}$}, Charles Tuffile\textit{$^{b}$}, Nicola Molinari,\textit{$^{b}$} Daniil Kitchaev,\textit{$^{b}$} Mordechai Kornbluth\textit{$^{b}$}} \\

\includegraphics{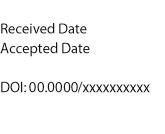} & \noindent\normalsize{
Polymer composite performance depends significantly on the polymer matrix, additives, processing conditions, and measurement setups. Traditional physics-based optimization methods for these parameters can be slow, labor-intensive, and costly, as they require physical manufacturing and testing. Here, we introduce a first step in extending Polymer Informatics, an AI-based approach proven effective for neat polymer design, into the realm of polymer composites. We curate a comprehensive database of commercially available polymer composites, develop a scheme for machine-readable data representation, and train machine-learning models for 15 flame-resistant, mechanical, thermal, and electrical properties, validating them on entirely unseen data. Future advancements are planned to drive the AI-assisted design of functional and sustainable polymer composites.
} \\

\end{tabular}

 \end{@twocolumnfalse} \vspace{0.6cm}

  ]

\renewcommand*\rmdefault{bch}\normalfont\upshape
\rmfamily
\section*{}
\vspace{-1cm}


\footnotetext{\textit{$^{a}$~Matmerize Inc., Atlanta, GA 30332, USA. E-mail: huan.tran@matmerize.com.}}
\footnotetext{\textit{$^{b}$~Robert Bosch LLC, Cambridge, MA 02139, USA.}}

\footnotetext{\dag~Electronic Supplementary Information (ESI) available: [details of any supplementary information available should be included here]. See DOI: 00.0000/00000000.}



\section{Introduction}\label{sec:intro}
\begin{figure}[t]
\centering
\includegraphics[width=0.42\textwidth]{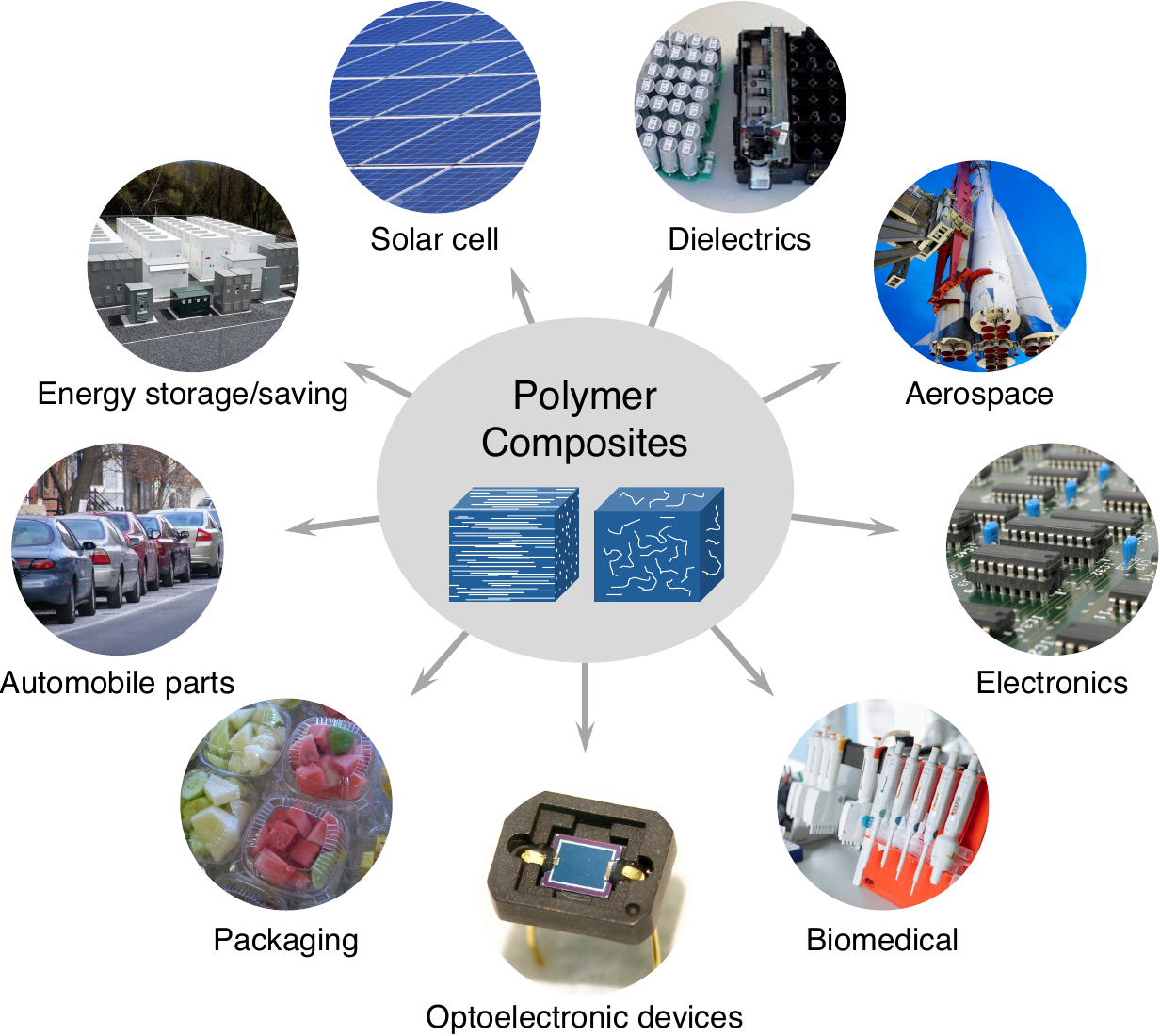}
	\caption{(Center panel) polymer composites, formed by implanting reinforcement fibers, fillers, or functional additives in a polymer matrix, and (surounding panels) their applications in different sectors of human life.}\label{fig:composite}
\end{figure}

Composites are materials created by combining two or more physically and chemically distinct phases to achieve desired properties or performance enhancements.\cite{barbero2010introduction, tong20023d, vilgis2009reinforcement} In polymer composites, as illustrated in Fig. \ref{fig:composite}, the main constituent phases include a matrix of base polymer, co-polymer, or polymer blend, and additional components such as reinforcement fibers, fillers, flame retardants, or functional additives.\cite{astrom2018manufacturing, yuan2019polymeric,vilgis2009reinforcement, tong20023d, irving2019polymer} A natural example of a polymer composite is wood, containing cellulose fibers embedded in lignin, a complex organic polymer.\cite{sarkanen1971lignins, ralph2004lignins} In this arrangement, the continuous lignin matrix carries and distributes applied loads among the cellulose fibers, giving wood its mechanical strength. By combining diverse polymer matrices, reinforcement fibers, and functional additives\cite{yuan2019polymeric,xie2018tune, he2021recent, zhang2020three, hu2017polymer, liu2021additive, thomas2012polymer, friedrich2005polymer, vilgis2009reinforcement, baillie2004green, maiti2022sustainable, islam2022graphene, hsissou2021polymer, kangishwar2023comprehensive} synthetic polymer composites may simultaneously meet multiple application-specific requirements, such as lightweight, high strength, corrosion resistance, durability under extreme conditions, and cost-effectiveness. As highlighted in Fig. \ref{fig:composite}, synthetic polymer composites are widely utilized across industries, including aerospace,\cite{irving2019polymer} automotive,\cite{elmarakbi2013advanced} and energy storage and conversion.\cite{grundish2021designing,zhu2021rational}

Designing polymer composites, i.e., rationally identifying formulations that meet predefined criteria for specific applications, is traditionally challenging, costly, and time-intensive, as candidates must be physically synthesized and tested.\cite{barbero2010introduction} Because of the inherent complexity of these materials, physics-based evaluation methods like molecular dynamics simulations and finite-element analysis are highly intricate, while quantum mechanical approaches such as density functional theory remain largely out of reach. Empirical models and rules, such as the ``rule of mixtures,''\cite{chawla1974applicability, tham2019tensile, hart1992ten,hart2002expanding} the ``Cox-Merz rule,''\cite{cox1958correlation} and the ``Halpin-Tsai equations,''\cite{affdl1976halpin,zhou2024modified} provide practical alternatives within specific domains but come with their own limitations.\cite{song2016empirical} Accordingly, a new robust evaluation method is essential to support experiments in polymer composite design.

\begin{figure*}[t]
\centering
\includegraphics[width=0.95\textwidth]{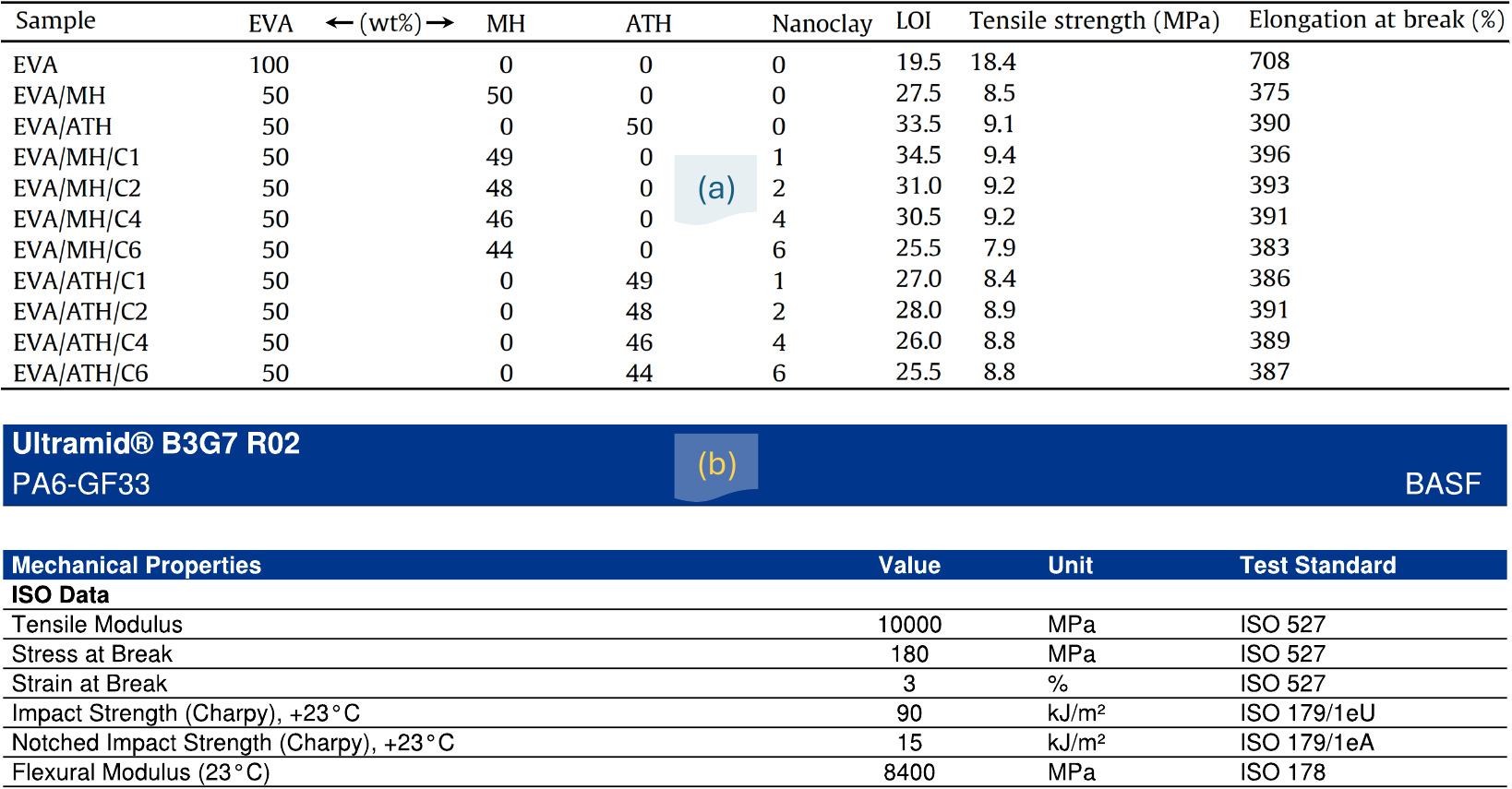}
	\caption{Two sources of polymer composites data curated for this work are (a) research articles and (b) technical datasheets/brochures provided by the manufacturers/distributors of commercialized products. Panel (a) was adapted from Ref. \citenum{yen2012synergistic} with permission while panel (b) was taken from a product brochure obtained from \texttt{https://www.albis.com}.}\label{fig:data}
\end{figure*}

Since the 2010s, machine-learning (ML) techniques have emerged as valuable complements to traditional approaches in materials science.\cite{Rampi:ML, Pilania_SR, Huan:design, Arun:design,Chiho:PG,doan2020machine,chen2021polymer, tran2024polymer,gurnani2024ai} In the field of polymer composites, ML has been used to accelerate simulations\cite{cassola2022machine} and predict physical properties\cite{sharma2022advances} such as conductivity,\cite{matos2019application,matos2019predictions} tensile strength,\cite{altarazi2018artificial,reddy2020modeling,zeng2019bp} fracture behavior,\cite{kushvaha2020artificial} and ductility.\cite{altarazi2018artificial,reddy2020modeling} The training data of these models are predominantly experimental in nature,\cite{altarazi2018artificial,kushvaha2020artificial,reddy2020modeling} while some of them were generated using finite element method.\cite{matos2019application,matos2019predictions} The data volume is typically small, ranging from less than ten\cite{zeng2019bp} to a few dozens,\cite{altarazi2018artificial,reddy2020modeling, ataeefard2019intelligently} and up to a few hundreds at most.\cite{kushvaha2020artificial} Apparently, data shortage is a major challenges in the future of accelerated design of polymeric materials.\cite{tran2024polymer}

This work aims to develop a set of robust ML models for polymer composites. To this end, we compiled and curated a database of over 5,000 polymer composites, fabricated in laboratories and/or industry, with multiple measured properties. Using this database, six multi-task ML models were trained and deployed to predict 15 properties in 4 groups, including flame resistance, mechanical, thermal, and electrical characteristics. The developed models demonstrate good performance on the validation data curated separately and kept unseen to the entire process. We believe that ML, when combined with sufficiently large and diverse datasets and suitable representations, offers a pathway toward the accelerated design of polymer composites.

\section{ML models for polymer composites}
\subsection{Data survey and curation}\label{sec:data}
Polymer composites are manufactured by carefully controlling the selection of the polymer matrix, additives, their compositions, and processing conditions. However, recorded information is often incomplete. Data on polymer composites typically comes from two main sources: research articles and technical datasheets or brochures from manufacturers. Generally, research articles supply more detailed information than technical datasheets. As illustrated in Fig. \ref{fig:data}(a), one study\cite{yen2012synergistic} examined composites with 50\% ethylene-vinyl acetate (EVA) matrix combined with some specific compositions of magnesium hydroxide (MH), aluminum trihydroxide (ATH), and nanoclay (modified montmorillonite), all considered potential flame retardants. Manufacturing, processing, and measurement details, along with measured flammability-related characteristics, can be found in this reference.\cite{yen2012synergistic}

Data provided in technical datasheets of commercialized polymer composites are generally less detailed. Fig. \ref{fig:data}(b) shows a top part of the brochure of ``Ultramid\textsuperscript{\textregistered} B3G7 R02'', a product of \textit{BASF}. This material is labeled as \texttt{PA6-GF33}, implying that it consists of Nylon 6 (PA6) as the polymer matrix and 33\% of glass fibers (GF). Such conventions are fairly standard across the polymer composite industry,\cite{elmarakbi2013advanced} although interpretations are not always straightforward. In case of ``ALCOM\textsuperscript{\textregistered} PA66 910/1.3 CF/GF30'', a product of \textit{MOCOM Compounds Corporation}, the label \texttt{PA66-(CF+GF)30} implies that it contains PA66 polymer matrix and a total of 30\% of glass fibers and carbon fibers (CF), but their separate compositions are unknown. Likewise, in the label of \texttt{(ABS+PA6)-GF8} used for ``Terblend\textsuperscript{\textregistered} N NG-02 EF'' (supplied by \textit{INEOS Styrolution}), the polymer matrix is a blend of \texttt{ABS} and \texttt{PA6}, but their compositions are also unavailable. In another example, the label of \texttt{PA6-GF30 FR} used for ``ALTECH PA6 A 2030/140 GF30 FR'' (also provided by \textit{MOCOM}) indicates that this material contains some flame retardants (FR), but does not provide their identity and compositions.

Our polymer composite database, curated from the two major sources and summarized in Table \ref{table:data}, contains 15 datasets for 15 flame-resistant, mechanical, thermal, and electrical properties. The flame-resistant datasets were curated from hundreds of research articles while the mechanical, thermal, and electrical datasets were extracted from about 10,000 technical datasheets, manually collected for about 5,000 commercialized polymer composites. The reported properties were measured under some widely recognized standards, e.g., ASTM E1354 (Cone calorimeter) and ASTM E662 (smoke chamber) for the flammability properties and ISO 527-1/-2 for the mechanical properties. 

\begin{figure}[t]
\centering
\includegraphics[width=0.475\textwidth]{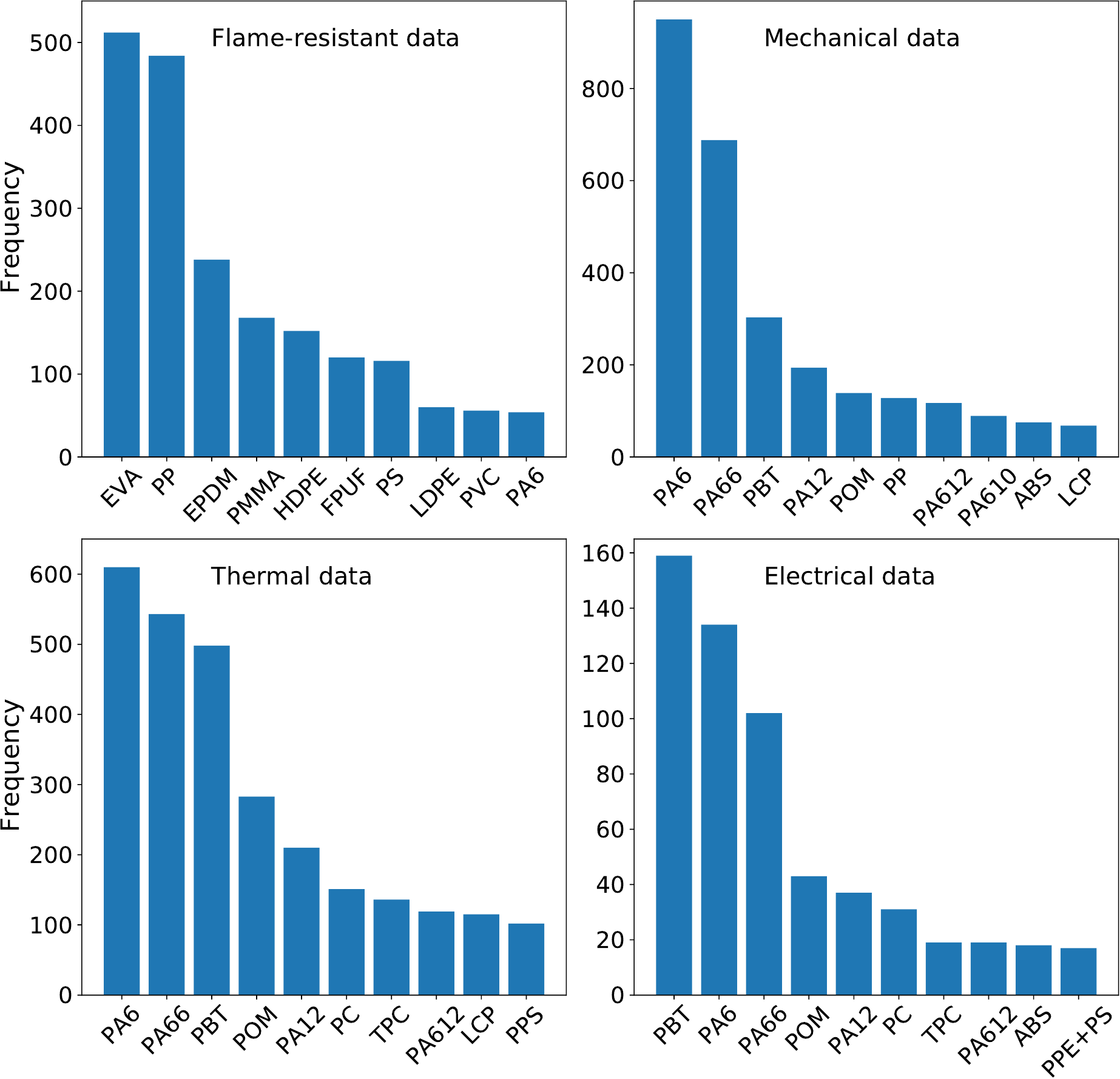}
\caption{Top ten base polymer matrices in four group of polymer composite datasets curated and used for this work.}\label{fig:stat}
\end{figure}

As discussed above, the description of the materials, needed for the inputs of the ML models, is generally more complete in the research articles than in technical datasheets. The identity and the composition of the polymer matrix and additives are available in the flame-resistant datasets. However, such information is not always available in the mechanical, thermal, and electrical datasets. In some entries, the compositions of polymer matrix blend and the additives may be missing. Notably, for those involving flame retardants, no information on their identity and composition is available. A snapshot of the flame-resistant, mechanical, thermal, and electrical datasets is given in Fig. \ref{fig:stat} while more information on the polymer matrices, the additives, and the flame retardants can be found in Supporting Information.

\begin{table}[t]
\caption{Summary of the datasets, including tensile modulus $E$, stress at break $\sigma_{\rm break}$, glass transition temperature $T_{\rm g}$, melting temperature $T_{\rm m}$, longitudinal coefficient of thermal expansion $\alpha_{\rm long}$, transverse coefficient of thermal expansion $\alpha_{\rm tran}$, relative permittivity at 1 MHz $\epsilon_{\rm 1 MHz}$, relative permittivity at 100 Hz $\epsilon_{\rm 100 Hz}$, the breakdown electric strength $E_{\rm bd}$, time to ignition TTI, peak heat release rate PHRR, averaged heat release rate AHRR, total heat release THR, optical smoke density $D_{\rm s}$, and maximum optical smoke density $D_{\rm max}$, curated and used in this work.}
    \centering
    \begin{tabular}{p{1.35cm} p{1.2cm} p{2.2cm} p{1.0cm} p{1.65cm} p{0.7cm}}
    \hline
	    \multirow{2}{*}{Class}&\multirow{2}{*}{Property} & \multirow{2}{*}{Standard}& \multirow{2}{*}{Unit} & \multirow{1}{*}{Data} & \multirow{1}{*}{Data}\\
	    &&&&\multirow{1}{*}{range}&\multirow{1}{*}{size}\\
            \hline
	                             & TTI            & ASTM E1354      & s           & 3.0 - 281.3& 527 \\
				     & PHRR                     & ASTM E1354      & kW/m$^2$     & 12.9 - 1876& 576\\
	    Flame                    & AHRR                     & ASTM E1354      & kW/m$^2$     & 58 - 750& 100 \\
	    resistant                & THR                      & ASTM E1354      & MJ/m$^2$     & 2.5 - 609& 316\\
                                     & $D_{\rm s}$              & ASTM E662       & --           & 0.1 - 857& 474\\
                                     & $D_{\rm max}$            & ASTM E662       & --           & 1.0 - 964& 124\\
        \hline                                                                                  
            Mechan-                     & $E$                      & ISO 527-1/-2    & MPa         & 7.4 - 38100&  4098 \\
            ical        & $\sigma_{\rm break}$     & ISO 527-1/-2    & MPa         & 12    - 329&  2738 \\
            \hline                                                                
	    \multirow{4}{*}{Thermal} & $T_{\rm g}$              & ISO 11357-1/-2  & C           & -109 - 337&  608 \\
	                             & $T_{\rm m}$              & ISO 11357-1/-3  & C           & $122$ - $388$&2044 \\
				     & $\alpha_{\rm long}$      & ISO 11359-1/-2  & $10^{-6}$/K & -2.4 - 250&3373 \\
				     & $\alpha_{\rm tran}$      & ISO 11359-1/-2  & $10^{-6}$/K & 1.17 - 230&2889\\
            \hline
            \multirow{3}{*}{Electrical}                         & $\epsilon_{\rm 100 Hz}$ & IEC 62631-2-1   & -- &2.5 - 15.0 &813\\
              & $\epsilon_{\rm 1 MHz}$  & IEC 62631-2-1   & -- &2.5 - 7.0&797 \\
                                         & $E_{\rm bd}$            & IEC 60243-1     & kV/mm &15 - 50&611 \\
         \hline
    \end{tabular}
    \label{table:data}
\end{table}

\begin{table*}[t]
    \centering
    \caption{Features used to develop the ML models.}\label{tab:feature}
	\begin{tabular}{p{3.5cm} p{9.0cm} p{5.8cm}}
    \hline
		Feature & Description & Applicable to\\
        \hline
		\texttt{cat\_polym} & Categorical, \texttt{PA6}, \texttt{ABS}, \texttt{PBT}, etc. & all models\\
		\texttt{num\_gf} & Numerical, composition of glass fibers & all models\\
		\texttt{num\_cf} & Numerical, composition of carbon fibers & all models\\
		\texttt{num\_gb} & Numerical, composition of glass beads & thermal, mechanical, \& electrical models\\
		\texttt{num\_md} & Numerical, composition of minerals& thermal, mechanical, \& electrical models\\
		\texttt{num\_density} & Numerical, material density (g/cm$^3$)& thermal, mechanical, \& electrical models\\
		\texttt{cat\_impact} & Categorical, Yes/No, if impact modifier included or not & thermal, mechanical, \& electrical models\\
		\texttt{cat\_condition} & Categorical, Dry/Conditioned, measurement condition & thermal, mechanical, \& electrical models\\
		\texttt{cat\_rif1 - cat\_rif2} & Categorical, identity of other reinforcements if included & flame-resistant models\\
		\texttt{num\_rif1 - num\_rif2} & Numerical, composition of other reinforcements if included & flame-resistant models\\
		\texttt{cat\_adv1 - cat\_adv2} & Categorical, identity of other additives if included & flame-resistant models\\
		\texttt{num\_adv1 - num\_adv2} & Numerical, composition of other additives if included & flame-resistant models\\
		\multirow{2}{*}{\texttt{cat\_fr1}} & Categorical, Yes/No, if first flame retardant included & thermal, mechanical, \& electrical models\\
		 & Categorical, identity of first flame retardant if included & flame-resistant models\\
		\texttt{num\_fr1} & Numerical, composition of first flame retardant & flame-resistant models\\
		\texttt{cat\_fr2 - cat\_fr4} & Categorical, identity of other flame retardants if included & flame-resistant models\\
		\texttt{num\_fr2 - num\_fr4} & Numerical, composition of other flame retardants & flame-resistant models\\
		\texttt{num\_cone\_heatflux} & Numerical, incoming heat flux (kW/m$^2$) in ASTM E1354 test& TTI, PHRR, AHRR, \& THR models\\
		\texttt{num\_cone\_thickness} & Numerical, thickness (mm) of the sample in ASTM E1354 test& TTI, PHRR, AHRR, \& THR models\\
		\texttt{num\_smoke\_heatflux} & Numerical, incoming heat flux (kW/m$^2$) in ASTM E662 test& $D_{\rm s}$ \& $D_{\rm max}$ models\\
		\texttt{num\_smoke\_thickness} & Numerical, thickness (mm) of the sample in ASTM E662 test& $D_{\rm s}$ \& $D_{\rm max}$ models\\
		\texttt{cat\_flaming} & Categorical, True/False, flaming mode in ASTM E662 test & $D_{\rm s}$ \& $D_{\rm max}$ models\\
		\texttt{num\_smoke\_time} & Numerical, time (s) of the optical smoke density measurement& $D_{\rm s}$ \& $D_{\rm max}$ models\\
	\hline
    \end{tabular}
\end{table*}

\begin{figure*}[t]
\centering
\includegraphics[width=1.0\textwidth]{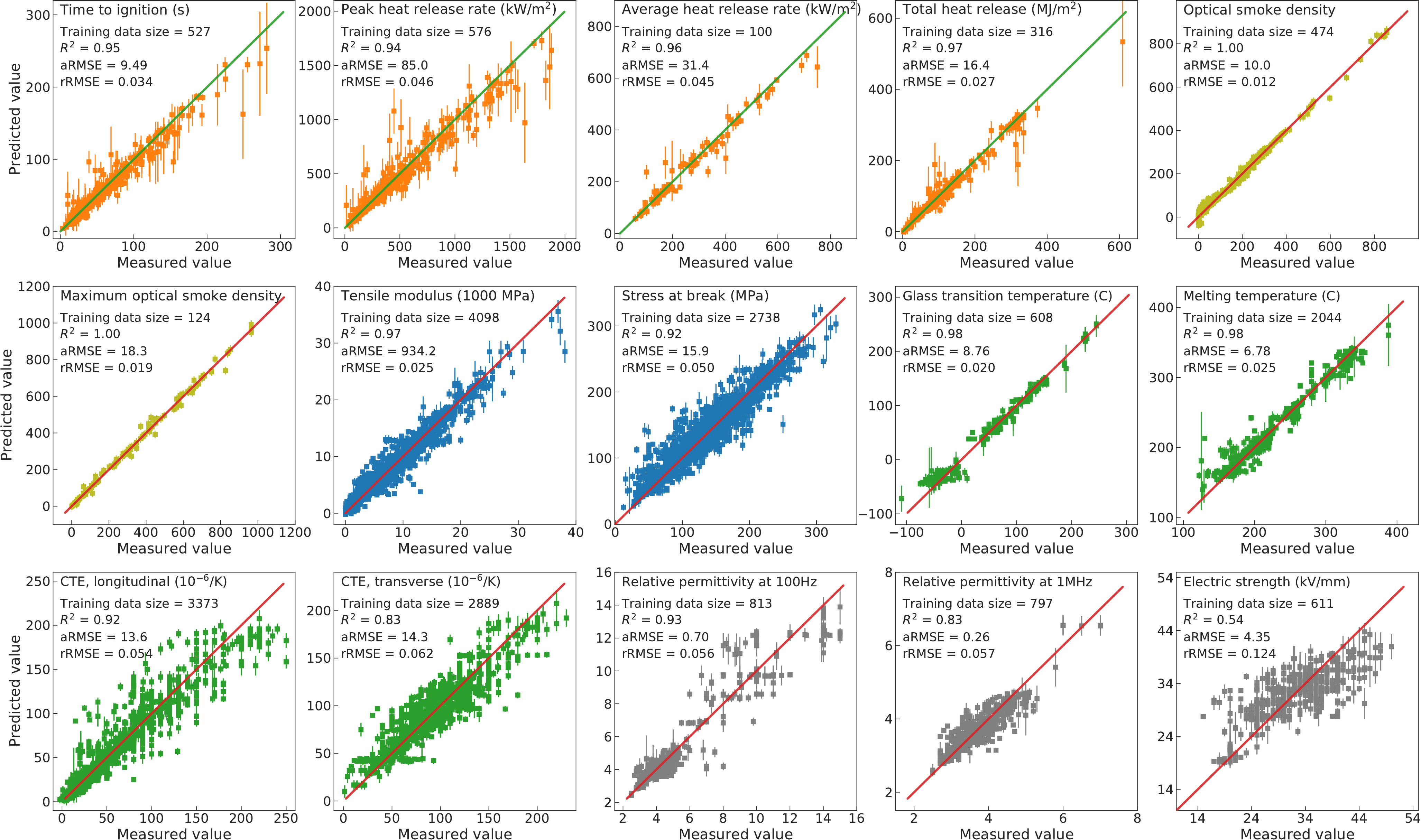}
	\caption{Visualization of 5 physics-informed MT models developed (and deployed in PolymRize\texttrademark) for 15 properties of polymer composites. For each of them, $R^2$ and aRMSE are provided. Each of 5 physics-informed MT models is marked by a distinct color.}\label{fig:models}
\end{figure*}

\subsection{Descriptors}\label{sec:feature}
Table \ref{tab:feature} summarizes the descriptors used to develop the models. Ideally, if SMILES strings\cite{smiles} encoding the chemical structure of polymer repeat units are available, they can be converted into numerical descriptors.\cite{Chiho:PG, doan2020machine, chen2021polymer} However, many polymer matrices in our database lack well-defined SMILES strings, as they are often cross-linking and/or without sufficient information, making descriptor calculation infeasible. Therefore, the polymer matrices are represented by a categorical descriptor, \texttt{cat\_polym}, taking their name (e.g., PA6, ABS, PBT) as its value. The composition of glass fibers, carbon fibers, glass beads, and minerals is captured by the numerical descriptors \texttt{num\_gf}, \texttt{num\_cf}, \texttt{num\_gb}, and \texttt{num\_md}, respectively. The presence of impact modifiers and flame retardants is indicated by \texttt{cat\_impact} and \texttt{cat\_fr1} (Yes/No values). Next, \texttt{cat\_condition} specifies the sample state during standard tests as either dry (fully dried) or conditioned (ambient exposure at 23°C and 50\% humidity). Lastly, due to missing details in the thermal, mechanical, and electrical datasets, the density (\texttt{num\_density}) is included as an augmentative descriptor, as it is consistently available across materials.

The flame-resistant models share several descriptors with the thermal, mechanical, and electrical models, including \texttt{cat\_polym}, \texttt{num\_gf}, \texttt{num\_cf}, and \texttt{cat\_fr1}. For \texttt{cat\_fr1} specifically, this descriptor specifies the identity of the first flame retardant, if present, while \texttt{num\_fr1} gives its composition. This numerical descriptor is unique to the flame-resistant models due to the absence of such data, as discussed above, in models of the other properties. Since materials in the flame-resistant datasets can contain up to four flame retardants, additional descriptors (\texttt{cat\_fr2}, \texttt{num\_fr2}, \texttt{cat\_fr3}, \texttt{num\_fr3}, \texttt{cat\_fr4}, \texttt{num\_fr4}) were included. Similarly, to account for up to two additional reinforcements and two additives beyond glass and carbon fibers, the descriptors \texttt{cat\_rif1}, \texttt{num\_rif1}, \texttt{cat\_rif2}, \texttt{num\_rif2}, \texttt{cat\_adv1}, \texttt{num\_adv1}, \texttt{cat\_adv2}, and \texttt{num\_adv2} were used.

Beyond material descriptors, additional features are required for the specific tests measuring flame-resistant performances. Cone calorimeter tests, conducted under ASTM E1354, measure time to ignition (TTI), peak heat release rate (PHRR), average heat release rate (AHRR), and total heat release (THR). Two key parameters of the tests, i.e., the incoming heat flux and the sample thickness, are described by \texttt{num\_cone\_heatflux} and \texttt{num\_cone\_thickness}. Likewise, smoke chamber tests, following ASTM E662, measure optical smoke density ($D_{\rm s}$) and maximum optical smoke density ($D_{\rm max}$) under flaming or non-flaming mode. Therefore, for $D_{\rm s}$ and $D_{\rm max}$ models, \texttt{num\_smoke\_heatflux}, \texttt{num\_smoke\_thickness}, \texttt{cat\_flaming} (flaming vs. non-flaming mode), and \texttt{num\_smoke\_time} (measurement time) are included, as $D_{\rm s}$ is time-dependent.

\begin{table}[t]
	\caption{Summary of five physics-informed MT models (separated by horizontal lines) developed for (1) time to ignition TTI, peak heat release rate PHRR, averaged heat release rate AHRR, and total heat release THR, (2) optical smoke density $D_{\rm s}$ and maximum optical smoke density $D_{\rm max}$, (3) tensile modulus $E$ and stress at break $\sigma_{\rm break}$, (4) glass transition temperature $T_{\rm g}$, melting temperature $T_{\rm m}$, longitudinal coefficient of thermal expansion $\alpha_{\rm long}$, and transverse coefficient of thermal expansion $\alpha_{\rm tran}$, and (5) relative permittivity at 1 MHz $\epsilon_{\rm 1 MHz}$, relative permittivity at 100 Hz $\epsilon_{\rm 100 Hz}$, and breakdown electric strength $E_{\rm bd}$.}
    \centering
	\begin{tabular}{p{0.94cm} p{1.65cm} p{0.75cm} p{0.95cm} p{0.9cm} p{0.01cm} p{0.75cm} p{.95cm} p{0.9cm}}
    \hline
		\multirow{2}{*}{Model} & \multirow{2}{*}{Algorithm}  & \multicolumn{3}{c}{Training} && \multicolumn{3}{c}{Validation} \\
            \cline{3-5} \cline{7-9} 
		&& $R^2$ & aRMSE & rRMSE && $R^2$ & aRMSE & rRMSE\\
            \hline
		TTI       	     	 &DL, Pi/MT& 0.95 & 9.5 &0.034&& NA & NA & NA\\
		PHRR                     &DL, Pi/MT& 0.94 & 85.0 &0.046&& NA & NA & NA\\
		AHRR                     &DL, Pi/MT& 0.96 & 31.4 &0.045&& NA & NA & NA \\
		THR                      &DL, Pi/MT& 0.97 & 16.4 &0.027&& NA & NA & NA\\
		\hline
		$D_{\rm s}$              &GPR, Pi/MT& 1.00 &10.1 & 0.012 && NA & NA & NA\\
		$D_{\rm max}$            &GPR, Pi/MT&1.00&18.3 & 0.019 && NA & NA & NA\\
            \hline                      
		$E$                      &DL, Pi/MT& 0.97 & 934 &0.025 &&0.98 & 624 & 0.030\\
		$\sigma_{\rm break}$     &DL, Pi/MT& 0.92 & 15.9 & 0.050 && 0.92 & 14.0 & 0.065\\
            \hline                   
		$T_{\rm g}$              &DL, Pi/MT& 0.98 & 8.76 & 0.020 && 0.97 & 8.54 & 0.038\\
		$T_{\rm m}$              &DL, Pi/MT& 0.98 & 6.78 & 0.025 && 0.98 & 3.13 &0.033\\
		$\alpha_{\rm long}$      &DL, Pi/MT& 0.92 & 13.6 & 0.054 && 0.92 & 11.2 &0.064\\
		$\alpha_{\rm tran}$      &DL, Pi/MT& 0.83 & 14.3 & 0.062  && 0.52& 13.7 & 0.121 \\
            \hline                    
		$\epsilon_{\rm 100 Hz}$  &DL, Pi/MT& 0.93 & 0.70 & 0.056 && NA & NA & NA\\
		$\epsilon_{\rm 1 MHz}$   &DL, Pi/MT& 0.83 & 0.26 & 0.057 && NA & NA & NA\\
		$E_{\rm bd}$             &DL, Pi/MT& 0.54 & 4.35 & 0.124 && NA & NA & NA\\
            \hline
    \end{tabular}
    \label{table:model}
\end{table}

\subsection{Model development and deployment}\label{sec:modeldev}
Two algorithms tested in this work are Gaussian process regression (GPR)\cite{GPRBook,GPR95} and deep learning (DL), both coupled with $k$-fold cross-validation. GPR, a similarity-based method, is intuitive and transparent. By modeling outputs as a Gaussian process, it provides prediction uncertainties that indicate whether a domain is well-represented in the training data. The DL models use a multilayer perceptron architecture with Leaky ReLU activation, Xavier-initialized dense layers,\cite{glorot2010understanding} batch normalization,\cite{Ioffe2015BatchShift} dropout,\cite{hinton2012improving} and ``UOut'' layers.\cite{Li2018UnderstandingShift} Xavier initialization prevents vanishing gradients, batch normalization normalizes layer inputs to reduce covariate shifts, and dropout mitigates overfitting by randomly deactivating units during training. The UOut layer resolves potential conflicts between dropout and batch normalization.

The DL models training uses Adam,\cite{Kingma2014Adam:Optimization} a gradient-based optimizer, to adjust parameters such as weights and biases, while hyperparameters are tuned using the following procedure. First, the number of layers (depth) that overfits the training data is identified. Then, key hyperparameters, i.e., learning rate, weight decay, batch size, and dropout percentage, are optimized. Afterward, k-fold cross-validation ensembles $k$ models, one per fold. This combination of dropout and cross-validation creates an ensemble of ensembles, enhancing predictive accuracy on unseen data. Prediction confidence is provided by computing the uncertainty $u$ for each test-time prediction $p_i$ from model $i$ in the ensemble. The uncertainty is given by $u = \sqrt{\frac{1}{N} \sum_{i=1}^{N} (p_i - \langle p_i\rangle)^2}$, where $N$ is the number of models in the ensemble, and $\langle p_i\rangle= \frac{1}{N} \sum_{i=1}^{N} p_i$. This uncertainty may be useful in some scenarios, e.g., ranking polymers based on performance confidence.

Traditionally, each ML model is trained independently on a single dataset in a procedure known as single-task (ST) learning. On the other hand, multi-task (MT) learning combines multiple related datasets to train a single model, leveraging potential correlations among material properties rooted in physical and chemical laws. Technically, these datasets are stacked together and indicated using an additional selector vector appended to the standard descriptors. The combined dataset can be used for any learning algorithm. In this work, MT learning is referred to as ``physics-informed'' (Pi) learning, as it uses augmentation data to implicitly convey these physics-containing correlations without requiring explicit mathematical expressions. Pi/MT approach is different from ``physics-enforced'' learning methods, which rely on directly encoding the correlations, given in terms of specific mathematical expressions, into the model. This study examines the Pi/MT approach against traditional ST learning for developing the targeted ML models (see Sec. \ref{sec:MT} for details).

We used the GPR and DL modules implemented in PolymRize\texttrademark\cite{polymrize_url} to train 15 ST models and 5 Pi/MT models for 15 properties summarized in Table \ref{table:data}. Among them, each Pi/MT model was trained on a set of properties that are intuitively/clearly correlated. For example, among 6 flame-resistant properties, TTI, PHRR, AHRR, and THR are typically measured simultaneously using a Cone calorimeter, thus they are clearly related and should be combined in a Pi/MT model. Likewise, $D_{\rm s}$ and $D_{\rm max}$ are measured simultaneously using a smoke chamber, thus another Pi/MT predictive model was developed for them. Starting from similar rationale, 3 other Pi/MT models were developed for the mechanical, thermal, and electrical properties.

\subsection{Model performances and validations}
As expected, the physics-informed MT models are systematically better than the corresponding ST models in multiple measures of performances, including the determination coefficient $R^2$, the absolute root-mean-square error aRMSE, and the relative root-mean-square error rRMSE, defined as the ratio between aRMSE and the whole range of the true data. While aRMSE cannot be compared across different datasets and models, rRMSE is more reliable for this purpose. These 3 performance metrics, computed on the training data, are summarized in Table \ref{table:model}. Among 15 models, 12 of them reach $R^2 > 0.9$, while other 2 models have $R^2 > 0.8$; rRMSE metric for all of them is about $5-6$\% and below. The electric strength model has a moderate $R^2 = 0.54$ and rRMSE $\simeq 12$\%. This result is reasonable and promising, given that the electric strength is related to and governed by multiple physics-based processes, spanning over multiple length and time scales, and thus understanding it is highly challenging.\cite{Huan:review, Arun:review, Vinit_Nature} These 5 models, visualized in Fig. \ref{fig:models}, are available in PolymRize\texttrademark.

\begin{figure}[t]
\centering
\includegraphics[width=0.485\textwidth]{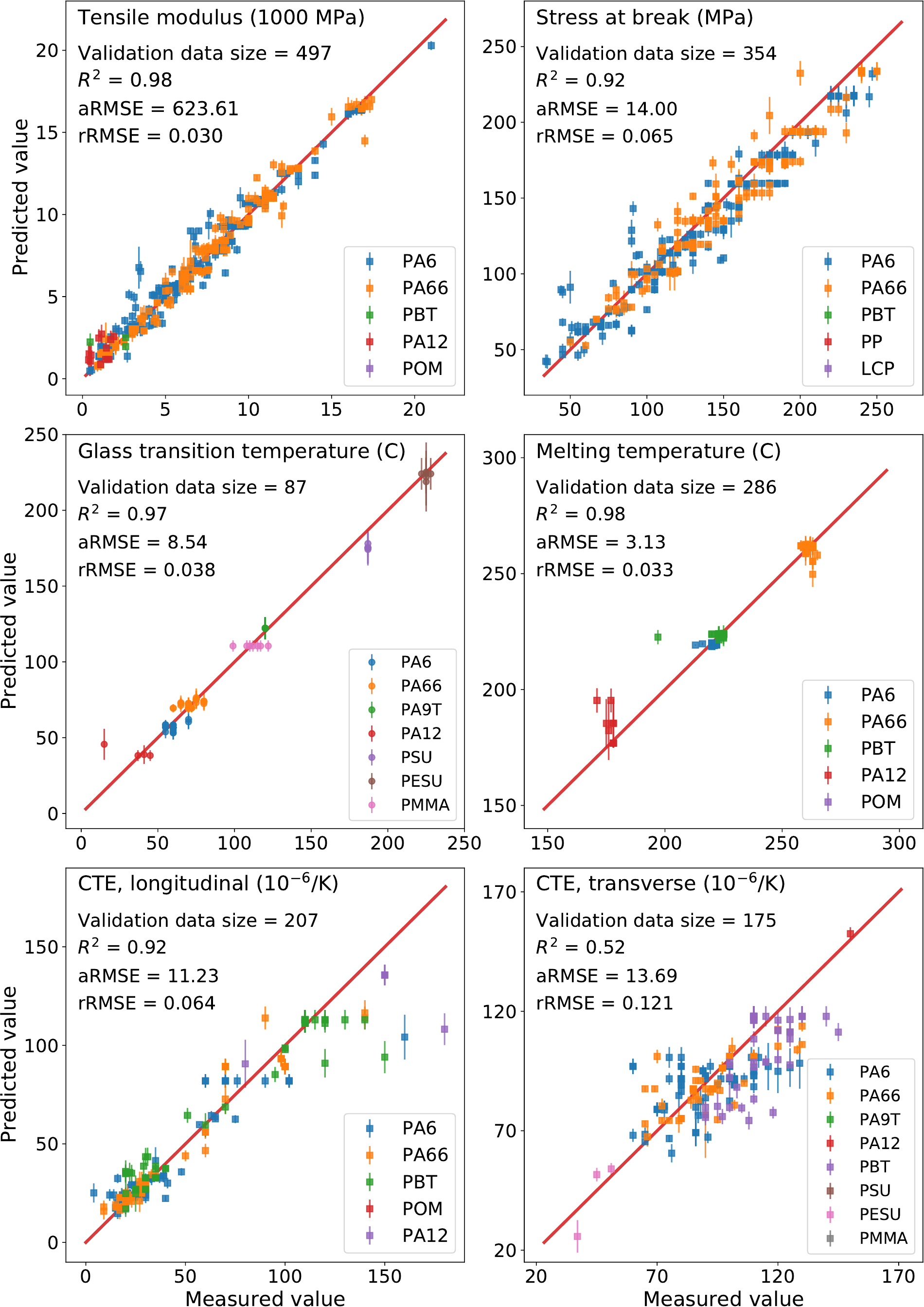}
\caption{Predictions of the deployed models for tensile modulus $E$, stress at break $\sigma_{\rm break}$, melting temperature $T_{\rm m}$, and longitudinal coefficient of thermal expansion $\alpha_{\rm long}$ on the unseen validation datasets curated completely independently. For each target property, the base polymer matrix of the validating materials are distinguished by colors.}\label{fig:valid}
\end{figure}

These deployed models were then validated on 6 completely unseen mechanical and thermal datasets curated independently by some team members, who have no access to the training data. For each target property, the validation data was then filtered, allowing only the composites having the base polymer matrices included in the list of 5 most-frequent matrices of the training data. Then, validation data were featurized and the targeted properties were predicted and compared with the ground truth. Predictions for tensile modulus $E$, stress at break $\sigma_{\rm break}$, glass transition temperature $T_{\rm g}$, melting temperature $T_{\rm m}$, longitudinal coefficient of thermal expansion $\alpha_{\rm long}$, and transverse coefficient of thermal expansion $\alpha_{\rm tran}$ on the unseen data are shown in Fig. \ref{fig:valid}. For all of the models except $\alpha_{\rm tran}$, the predictions agree very well with the ground truth with $R^2 > 90$\% and aRMSE that is comparable with that reported in Table \ref{table:model}. In summary, the mechanical and thermal properties models can predict the unseen data accurately, suggesting that the training data of these models are sufficiently big and diverse to represent the common cases of polymer composites.

\section{Physics-informed MT learning approach}\label{sec:MT}
\begin{figure}[t]
\centering
\includegraphics[width=0.475\textwidth]{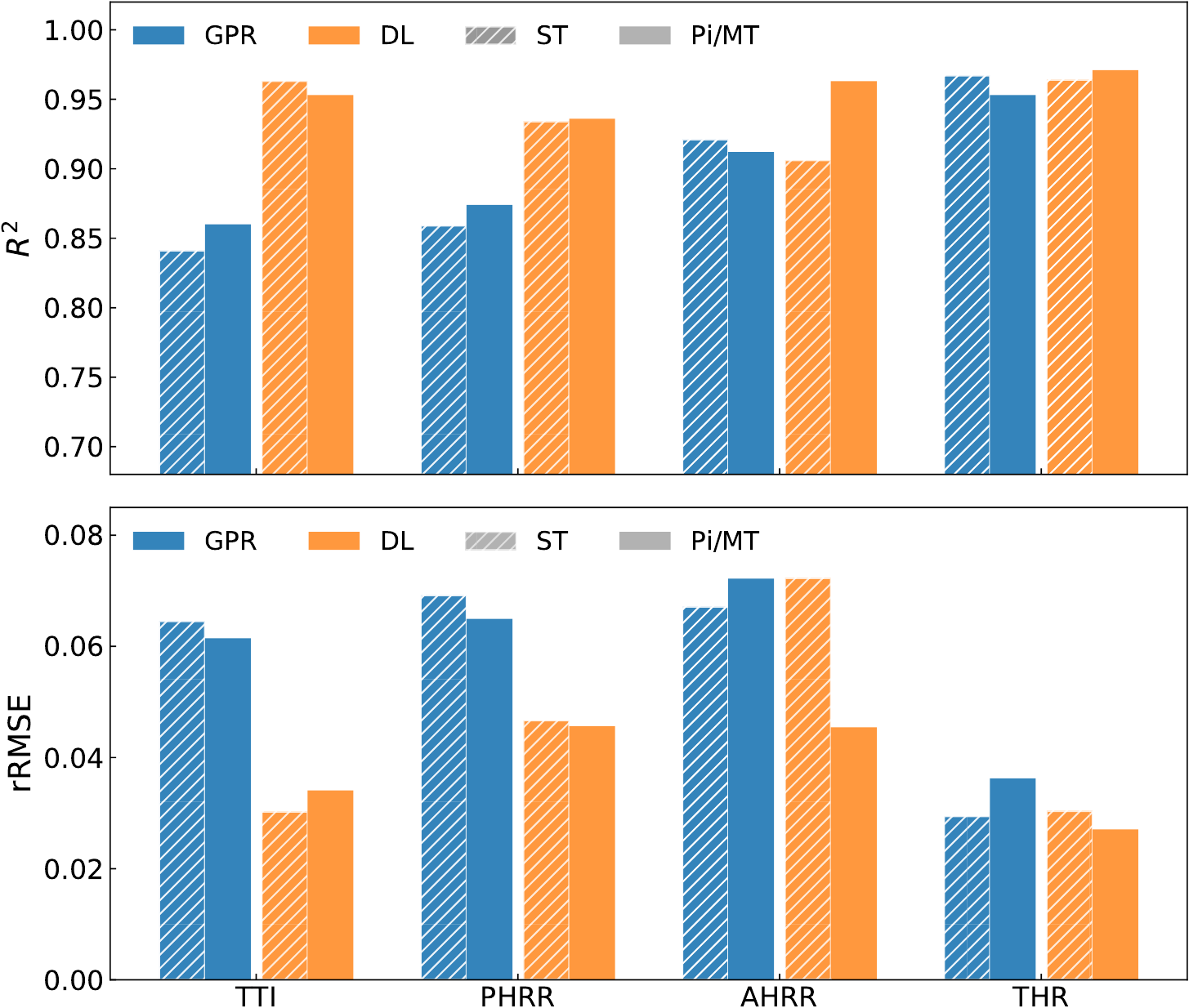}
\caption{Coefficient of determination $R^2$ and relative RMSE of the models trained for the time to ignition TTI, the peak heat release rate PHRR, the averaged heat release rate AHRR, and the total heat release THR. For both GPR and DT, ST and physics-informed MT models are indicated by crossed and solid patterns, respectively.}\label{fig:MT}
\end{figure}

The advantage of the physics-informed MT models over their ST counterparts is desirable and expected when the correlations among the training datasets are strong. Fig. \ref{fig:MT} provides a summary of the models trained for time to ignition TTI, peak heat release rate PHRR, averaged heat release rate AHRR, and total heat release THR. Between two learning algorithms, GPR is relatively less effective with TTI and PHRR than DL, for which $R^2$ is consistently higher than 90\%. The physics-informed MT model trained using DL is clearly better, leveraging $R^2$ of all four properties above 95\% while making them more comparable. Similarly, rRMSE is significantly reduced and becomes more balance across TTI, PHRR, AHRR, and THR. 

The main rationale of the Pi/MT approach is that by deliberately generating, producing, supplying, and thus, ``informing''  the training process with data of related properties, the target ML models can be improved.\cite{tran2024polymer} There is, in principle, no limit in the nature and the volume of the augmented data. Moreover, the expected correlations among the datasets are not required to be materialized into any solid mathematical expression. With these two major advantages, the physics-informed MT approach is expected to be widely used in the research area of polymer composites.\cite{tran2024polymer}

\section{Forward-looking perspectives and conclusions}
Polymer composite data are often scarce and incomplete, posing significant challenges for developing ML predictive models. Addressing these challenges could unlock opportunities for accelerated property predictions and polymer composite design, specifically for extremes. By deploying five Pi/MT models on the largest datasets of their kind and demonstrating predictive performance across 15 widely used properties, this work marks an initial step toward that future.

From the ML perspective, the physics-informed MT learning approach consistently outperformed traditional ST learning, where each model is independently developed for a single property. Prior studies\cite{kuenneth2021polymer, gurnani2023polymer} suggest that MT architectures can capture hidden correlations among related properties. This work supports that theory. Nevertheless, small data size and large data noise, both of which are common in practice, can suppress the correlations and limit the MT learning efficiency. Addressing these issues remains open for future works.

Manual data curation, as performed here, is unsustainable given the abundance of polymer composite data. Advances in natural language processing, including large language models, named entity recognition, normalization, relation extraction, and co-referencing, may soon offer scalable solutions. Additionally, representing base polymers by name or label, as done in this study, is suboptimal. Future improvements could involve acquiring SMILES strings\cite{smiles} for all polymers and extending chemical fingerprinting schemes\cite{Chiho:PG,doan2020machine} to better handle cross-linking polymers and other complex classes, further advancing model performance.

\section*{Data Availability}
Data supporting the discussions and conclusions of this work are available in the main text and the Supporting Information.

\section*{Author Contributions}
H.T. designed the project and wrote the manuscript. O.H., K.G., C.T., N.M., D.K., and M.K. collected, cleaned, and contributed to processing data, C.K., R.G., and J.D. implemented Pi/MT, deployed, and tested the models, R.R. helped with designing the project and writing the manuscript.

\section*{Conflicts of interest}
The authors declare no conflicts of interest.

\section*{Acknowledgements}
HT, RG, and OV acknowledge financial supports from National Science Foundation through the SBIR Phase I Grant \#2322108. HT thanks Office of Naval Research for financial supports through the SBIR Phase I Contract \#N68335-24-C-0121. The authors thank Sydney Balcom (Matmerize, Inc.) for technical assistance.



\balance

\renewcommand\refname{References}


\begin{mcitethebibliography}{64}
\providecommand*{\natexlab}[1]{#1}
\providecommand*{\mciteSetBstSublistMode}[1]{}
\providecommand*{\mciteSetBstMaxWidthForm}[2]{}
\providecommand*{\mciteBstWouldAddEndPuncttrue}
  {\def\EndOfBibitem{\unskip.}}
\providecommand*{\mciteBstWouldAddEndPunctfalse}
  {\let\EndOfBibitem\relax}
\providecommand*{\mciteSetBstMidEndSepPunct}[3]{}
\providecommand*{\mciteSetBstSublistLabelBeginEnd}[3]{}
\providecommand*{\EndOfBibitem}{}
\mciteSetBstSublistMode{f}
\mciteSetBstMaxWidthForm{subitem}
{(\emph{\alph{mcitesubitemcount}})}
\mciteSetBstSublistLabelBeginEnd{\mcitemaxwidthsubitemform\space}
{\relax}{\relax}

\bibitem[Barbero(2010)]{barbero2010introduction}
E.~J. Barbero, \emph{Introduction to composite materials design}, CRC press,
  2010\relax
\mciteBstWouldAddEndPuncttrue
\mciteSetBstMidEndSepPunct{\mcitedefaultmidpunct}
{\mcitedefaultendpunct}{\mcitedefaultseppunct}\relax
\EndOfBibitem
\bibitem[Tong \emph{et~al.}(2002)Tong, Mouritz, and Bannister]{tong20023d}
L.~Tong, A.~P. Mouritz and M.~Bannister, \emph{3D fibre reinforced polymer
  composites}, Elsevier, 2002\relax
\mciteBstWouldAddEndPuncttrue
\mciteSetBstMidEndSepPunct{\mcitedefaultmidpunct}
{\mcitedefaultendpunct}{\mcitedefaultseppunct}\relax
\EndOfBibitem
\bibitem[Vilgis \emph{et~al.}(2009)Vilgis, Heinrich, and
  Kl{\"u}ppel]{vilgis2009reinforcement}
T.~A. Vilgis, G.~Heinrich and M.~Kl{\"u}ppel, \emph{Reinforcement of polymer
  nano-composites: theory, experiments and applications}, Cambridge University
  Press, 2009\relax
\mciteBstWouldAddEndPuncttrue
\mciteSetBstMidEndSepPunct{\mcitedefaultmidpunct}
{\mcitedefaultendpunct}{\mcitedefaultseppunct}\relax
\EndOfBibitem
\bibitem[Astrom(2018)]{astrom2018manufacturing}
B.~T. Astrom, \emph{Manufacturing of polymer composites}, Routledge, 2018\relax
\mciteBstWouldAddEndPuncttrue
\mciteSetBstMidEndSepPunct{\mcitedefaultmidpunct}
{\mcitedefaultendpunct}{\mcitedefaultseppunct}\relax
\EndOfBibitem
\bibitem[Yuan \emph{et~al.}(2019)Yuan, Shen, Chua, and Zhou]{yuan2019polymeric}
S.~Yuan, F.~Shen, C.~K. Chua and K.~Zhou, \emph{Prog Polym. Sci.}, 2019,
  \textbf{91}, 141--168\relax
\mciteBstWouldAddEndPuncttrue
\mciteSetBstMidEndSepPunct{\mcitedefaultmidpunct}
{\mcitedefaultendpunct}{\mcitedefaultseppunct}\relax
\EndOfBibitem
\bibitem[Irving and Soutis(2019)]{irving2019polymer}
P.~E. Irving and C.~Soutis, \emph{Polymer composites in the aerospace
  industry}, Woodhead Publishing, 2019\relax
\mciteBstWouldAddEndPuncttrue
\mciteSetBstMidEndSepPunct{\mcitedefaultmidpunct}
{\mcitedefaultendpunct}{\mcitedefaultseppunct}\relax
\EndOfBibitem
\bibitem[Sarkanen and Ludwig(1971)]{sarkanen1971lignins}
K.~V. Sarkanen and C.~H. Ludwig, \emph{Lignins: occurrence, formation,
  structure and reactions}, Wiley-Interscience, New York, 1971\relax
\mciteBstWouldAddEndPuncttrue
\mciteSetBstMidEndSepPunct{\mcitedefaultmidpunct}
{\mcitedefaultendpunct}{\mcitedefaultseppunct}\relax
\EndOfBibitem
\bibitem[Ralph \emph{et~al.}(2004)Ralph, Lundquist, Brunow, Lu, Kim, Schatz,
  Marita, Hatfield, Ralph, Christensen,\emph{et~al.}]{ralph2004lignins}
J.~Ralph, K.~Lundquist, G.~Brunow, F.~Lu, H.~Kim, P.~F. Schatz, J.~M. Marita,
  R.~D. Hatfield, S.~A. Ralph, J.~H. Christensen \emph{et~al.},
  \emph{Phytochem. Rev.}, 2004, \textbf{3}, 29--60\relax
\mciteBstWouldAddEndPuncttrue
\mciteSetBstMidEndSepPunct{\mcitedefaultmidpunct}
{\mcitedefaultendpunct}{\mcitedefaultseppunct}\relax
\EndOfBibitem
\bibitem[Xie and Zhu(2018)]{xie2018tune}
L.~Xie and Y.~Zhu, \emph{Polym. Compos.}, 2018, \textbf{39}, 2985--2996\relax
\mciteBstWouldAddEndPuncttrue
\mciteSetBstMidEndSepPunct{\mcitedefaultmidpunct}
{\mcitedefaultendpunct}{\mcitedefaultseppunct}\relax
\EndOfBibitem
\bibitem[He and Wang(2021)]{he2021recent}
X.~He and Y.~Wang, \emph{Ind. Eng. Chem. Res.}, 2021, \textbf{60},
  1137--1154\relax
\mciteBstWouldAddEndPuncttrue
\mciteSetBstMidEndSepPunct{\mcitedefaultmidpunct}
{\mcitedefaultendpunct}{\mcitedefaultseppunct}\relax
\EndOfBibitem
\bibitem[Zhang \emph{et~al.}(2020)Zhang, Feng, and Feng]{zhang2020three}
F.~Zhang, Y.~Feng and W.~Feng, \emph{Mater. Sci. Eng. R Rep.}, 2020,
  \textbf{142}, 100580\relax
\mciteBstWouldAddEndPuncttrue
\mciteSetBstMidEndSepPunct{\mcitedefaultmidpunct}
{\mcitedefaultendpunct}{\mcitedefaultseppunct}\relax
\EndOfBibitem
\bibitem[Hu \emph{et~al.}(2017)Hu, Huang, Yao, Pan, Sun, Zeng, Sun, Xu, Song,
  and Wong]{hu2017polymer}
J.~Hu, Y.~Huang, Y.~Yao, G.~Pan, J.~Sun, X.~Zeng, R.~Sun, J.-B. Xu, B.~Song and
  C.-P. Wong, \emph{ACS Appl. Mater. Interfaces}, 2017, \textbf{9},
  13544--13553\relax
\mciteBstWouldAddEndPuncttrue
\mciteSetBstMidEndSepPunct{\mcitedefaultmidpunct}
{\mcitedefaultendpunct}{\mcitedefaultseppunct}\relax
\EndOfBibitem
\bibitem[Liu \emph{et~al.}(2021)Liu, Xiong, and Zhou]{liu2021additive}
G.~Liu, Y.~Xiong and L.~Zhou, \emph{Compos. Commun.}, 2021, \textbf{27},
  100907\relax
\mciteBstWouldAddEndPuncttrue
\mciteSetBstMidEndSepPunct{\mcitedefaultmidpunct}
{\mcitedefaultendpunct}{\mcitedefaultseppunct}\relax
\EndOfBibitem
\bibitem[Thomas \emph{et~al.}(2012)Thomas, Joseph, Malhotra, Goda, and
  Sreekala]{thomas2012polymer}
S.~Thomas, K.~Joseph, S.~K. Malhotra, K.~Goda and M.~S. Sreekala, \emph{Polymer
  composites, macro-and microcomposites}, John Wiley \& Sons, 2012,
  vol.~1\relax
\mciteBstWouldAddEndPuncttrue
\mciteSetBstMidEndSepPunct{\mcitedefaultmidpunct}
{\mcitedefaultendpunct}{\mcitedefaultseppunct}\relax
\EndOfBibitem
\bibitem[Friedrich \emph{et~al.}(2005)Friedrich, Fakirov, and
  Zhang]{friedrich2005polymer}
K.~Friedrich, S.~Fakirov and Z.~Zhang, \emph{Polymer composites: from nano-to
  macro-scale}, Springer Science \& Business Media, 2005\relax
\mciteBstWouldAddEndPuncttrue
\mciteSetBstMidEndSepPunct{\mcitedefaultmidpunct}
{\mcitedefaultendpunct}{\mcitedefaultseppunct}\relax
\EndOfBibitem
\bibitem[Baillie and Jayasinghe(2004)]{baillie2004green}
C.~Baillie and R.~Jayasinghe, \emph{Green composites: polymer composites and
  the environment}, Elsevier, 2004\relax
\mciteBstWouldAddEndPuncttrue
\mciteSetBstMidEndSepPunct{\mcitedefaultmidpunct}
{\mcitedefaultendpunct}{\mcitedefaultseppunct}\relax
\EndOfBibitem
\bibitem[Maiti \emph{et~al.}(2022)Maiti, Islam, Uddin, Afroj, Eichhorn, and
  Karim]{maiti2022sustainable}
S.~Maiti, M.~R. Islam, M.~A. Uddin, S.~Afroj, S.~J. Eichhorn and N.~Karim,
  \emph{Adv. Sustain. Syst.}, 2022, \textbf{6}, 2200258\relax
\mciteBstWouldAddEndPuncttrue
\mciteSetBstMidEndSepPunct{\mcitedefaultmidpunct}
{\mcitedefaultendpunct}{\mcitedefaultseppunct}\relax
\EndOfBibitem
\bibitem[Islam \emph{et~al.}(2022)Islam, Afroj, Uddin, Andreeva, Novoselov, and
  Karim]{islam2022graphene}
M.~H. Islam, S.~Afroj, M.~A. Uddin, D.~V. Andreeva, K.~S. Novoselov and
  N.~Karim, \emph{Adv. Funct. Mater.}, 2022, \textbf{32}, 2205723\relax
\mciteBstWouldAddEndPuncttrue
\mciteSetBstMidEndSepPunct{\mcitedefaultmidpunct}
{\mcitedefaultendpunct}{\mcitedefaultseppunct}\relax
\EndOfBibitem
\bibitem[Hsissou \emph{et~al.}(2021)Hsissou, Seghiri, Benzekri, Hilali, Rafik,
  and Elharfi]{hsissou2021polymer}
R.~Hsissou, R.~Seghiri, Z.~Benzekri, M.~Hilali, M.~Rafik and A.~Elharfi,
  \emph{Compos. Struct.}, 2021, \textbf{262}, 113640\relax
\mciteBstWouldAddEndPuncttrue
\mciteSetBstMidEndSepPunct{\mcitedefaultmidpunct}
{\mcitedefaultendpunct}{\mcitedefaultseppunct}\relax
\EndOfBibitem
\bibitem[Kangishwar \emph{et~al.}(2023)Kangishwar, Radhika, Sheik, Chavali, and
  Hariharan]{kangishwar2023comprehensive}
S.~Kangishwar, N.~Radhika, A.~A. Sheik, A.~Chavali and S.~Hariharan,
  \emph{Polym. Bull.}, 2023, \textbf{80}, 47--87\relax
\mciteBstWouldAddEndPuncttrue
\mciteSetBstMidEndSepPunct{\mcitedefaultmidpunct}
{\mcitedefaultendpunct}{\mcitedefaultseppunct}\relax
\EndOfBibitem
\bibitem[Elmarakbi(2013)]{elmarakbi2013advanced}
A.~Elmarakbi, \emph{Advanced composite materials for automotive applications:
  Structural integrity and crashworthiness}, John Wiley \& Sons, 2013\relax
\mciteBstWouldAddEndPuncttrue
\mciteSetBstMidEndSepPunct{\mcitedefaultmidpunct}
{\mcitedefaultendpunct}{\mcitedefaultseppunct}\relax
\EndOfBibitem
\bibitem[Grundish \emph{et~al.}(2021)Grundish, Goodenough, and
  Khani]{grundish2021designing}
N.~S. Grundish, J.~B. Goodenough and H.~Khani, \emph{Curr. Opin. Electrochem.},
  2021, \textbf{30}, 100828\relax
\mciteBstWouldAddEndPuncttrue
\mciteSetBstMidEndSepPunct{\mcitedefaultmidpunct}
{\mcitedefaultendpunct}{\mcitedefaultseppunct}\relax
\EndOfBibitem
\bibitem[Zhu \emph{et~al.}(2021)Zhu, Yu, Song, Chen, and Chen]{zhu2021rational}
M.-X. Zhu, Q.-C. Yu, H.-G. Song, T.-X. Chen and J.-M. Chen, \emph{ACS Appl.
  Energy Mater.}, 2021, \textbf{4}, 1449--1458\relax
\mciteBstWouldAddEndPuncttrue
\mciteSetBstMidEndSepPunct{\mcitedefaultmidpunct}
{\mcitedefaultendpunct}{\mcitedefaultseppunct}\relax
\EndOfBibitem
\bibitem[Chawla(1974)]{chawla1974applicability}
K.~Chawla, \emph{Rev. Bras. F{\'\i}s.}, 1974, \textbf{4}, 411--418\relax
\mciteBstWouldAddEndPuncttrue
\mciteSetBstMidEndSepPunct{\mcitedefaultmidpunct}
{\mcitedefaultendpunct}{\mcitedefaultseppunct}\relax
\EndOfBibitem
\bibitem[Tham \emph{et~al.}(2019)Tham, Fazita, Abdul~Khalil, Mahmud~Zuhudi,
  Jaafar, Rizal, and Haafiz]{tham2019tensile}
M.~W. Tham, M.~N. Fazita, H.~Abdul~Khalil, N.~Z. Mahmud~Zuhudi, M.~Jaafar,
  S.~Rizal and M.~M. Haafiz, \emph{J. Reinf. Plast. Compos.}, 2019,
  \textbf{38}, 211--248\relax
\mciteBstWouldAddEndPuncttrue
\mciteSetBstMidEndSepPunct{\mcitedefaultmidpunct}
{\mcitedefaultendpunct}{\mcitedefaultseppunct}\relax
\EndOfBibitem
\bibitem[Hart-Smith(1992)]{hart1992ten}
L.~Hart-Smith, \emph{Weight Engineering}, 1992, \textbf{52}, 29--45\relax
\mciteBstWouldAddEndPuncttrue
\mciteSetBstMidEndSepPunct{\mcitedefaultmidpunct}
{\mcitedefaultendpunct}{\mcitedefaultseppunct}\relax
\EndOfBibitem
\bibitem[Hart-Smith(2002)]{hart2002expanding}
L.~Hart-Smith, \emph{Compos. Sci. Technol.}, 2002, \textbf{62},
  1515--1544\relax
\mciteBstWouldAddEndPuncttrue
\mciteSetBstMidEndSepPunct{\mcitedefaultmidpunct}
{\mcitedefaultendpunct}{\mcitedefaultseppunct}\relax
\EndOfBibitem
\bibitem[Cox and Merz(1958)]{cox1958correlation}
W.~Cox and E.~Merz, \emph{J. Polym. Sci.}, 1958, \textbf{28}, 619--622\relax
\mciteBstWouldAddEndPuncttrue
\mciteSetBstMidEndSepPunct{\mcitedefaultmidpunct}
{\mcitedefaultendpunct}{\mcitedefaultseppunct}\relax
\EndOfBibitem
\bibitem[Affdl and Kardos(1976)]{affdl1976halpin}
J.~H. Affdl and J.~Kardos, \emph{Poly. Eng. Sci.}, 1976, \textbf{16},
  344--352\relax
\mciteBstWouldAddEndPuncttrue
\mciteSetBstMidEndSepPunct{\mcitedefaultmidpunct}
{\mcitedefaultendpunct}{\mcitedefaultseppunct}\relax
\EndOfBibitem
\bibitem[Zhou \emph{et~al.}(2024)Zhou, Tong, Liu, Lv, Srivatsan, and
  Gao]{zhou2024modified}
D.~Zhou, X.~Tong, H.~Liu, S.~Lv, T.~Srivatsan and X.~Gao, \emph{AIP Adv.},
  2024, \textbf{14}, year\relax
\mciteBstWouldAddEndPuncttrue
\mciteSetBstMidEndSepPunct{\mcitedefaultmidpunct}
{\mcitedefaultendpunct}{\mcitedefaultseppunct}\relax
\EndOfBibitem
\bibitem[Song \emph{et~al.}(2016)Song, Muliana, and
  Palazotto]{song2016empirical}
R.~Song, A.~H. Muliana and A.~Palazotto, \emph{Compos. Struct.}, 2016,
  \textbf{148}, 207--223\relax
\mciteBstWouldAddEndPuncttrue
\mciteSetBstMidEndSepPunct{\mcitedefaultmidpunct}
{\mcitedefaultendpunct}{\mcitedefaultseppunct}\relax
\EndOfBibitem
\bibitem[Yen \emph{et~al.}(2012)Yen, Wang, and Guo]{yen2012synergistic}
Y.-Y. Yen, H.-T. Wang and W.-J. Guo, \emph{Polym. Degrad. Stab.}, 2012,
  \textbf{97}, 863--869\relax
\mciteBstWouldAddEndPuncttrue
\mciteSetBstMidEndSepPunct{\mcitedefaultmidpunct}
{\mcitedefaultendpunct}{\mcitedefaultseppunct}\relax
\EndOfBibitem
\bibitem[Ramprasad \emph{et~al.}(2017)Ramprasad, Batra, Pilania,
  Mannodi-Kanakkithodi, and Kim]{Rampi:ML}
R.~Ramprasad, R.~Batra, G.~Pilania, A.~Mannodi-Kanakkithodi and C.~Kim,
  \emph{npj Comput. Mater.}, 2017, \textbf{3}, 54\relax
\mciteBstWouldAddEndPuncttrue
\mciteSetBstMidEndSepPunct{\mcitedefaultmidpunct}
{\mcitedefaultendpunct}{\mcitedefaultseppunct}\relax
\EndOfBibitem
\bibitem[Pilania \emph{et~al.}(2013)Pilania, Wang, Jiang, Rajasekaran, and
  Ramprasad]{Pilania_SR}
G.~Pilania, C.~Wang, X.~Jiang, S.~Rajasekaran and R.~Ramprasad, \emph{Sci.
  Rep.}, 2013, \textbf{3}, 2810\relax
\mciteBstWouldAddEndPuncttrue
\mciteSetBstMidEndSepPunct{\mcitedefaultmidpunct}
{\mcitedefaultendpunct}{\mcitedefaultseppunct}\relax
\EndOfBibitem
\bibitem[Huan \emph{et~al.}(2015)Huan, Mannodi-Kanakkithodi, and
  Ramprasad]{Huan:design}
T.~D. Huan, A.~Mannodi-Kanakkithodi and R.~Ramprasad, \emph{Phys. Rev. B},
  2015, \textbf{92}, 014106\relax
\mciteBstWouldAddEndPuncttrue
\mciteSetBstMidEndSepPunct{\mcitedefaultmidpunct}
{\mcitedefaultendpunct}{\mcitedefaultseppunct}\relax
\EndOfBibitem
\bibitem[Mannodi-Kanakkithodi \emph{et~al.}(2016)Mannodi-Kanakkithodi, Pilania,
  Huan, Lookman, and Ramprasad]{Arun:design}
A.~Mannodi-Kanakkithodi, G.~Pilania, T.~D. Huan, T.~Lookman and R.~Ramprasad,
  \emph{Sci. Rep.}, 2016, \textbf{6}, 20952\relax
\mciteBstWouldAddEndPuncttrue
\mciteSetBstMidEndSepPunct{\mcitedefaultmidpunct}
{\mcitedefaultendpunct}{\mcitedefaultseppunct}\relax
\EndOfBibitem
\bibitem[Kim \emph{et~al.}(2018)Kim, Chandrasekaran, Huan, Das, and
  Ramprasad]{Chiho:PG}
C.~Kim, A.~Chandrasekaran, T.~D. Huan, D.~Das and R.~Ramprasad, \emph{J. Phys.
  Chem. C}, 2018, \textbf{122}, 17575--17585\relax
\mciteBstWouldAddEndPuncttrue
\mciteSetBstMidEndSepPunct{\mcitedefaultmidpunct}
{\mcitedefaultendpunct}{\mcitedefaultseppunct}\relax
\EndOfBibitem
\bibitem[Tran \emph{et~al.}(2020)Tran, Kim, Chen, Chandrasekaran, Batra,
  Venkatram, Kamal, Lightstone, Gurnani, Shetty, Ramprasad, Laws, Shelton, and
  Ramprasad]{doan2020machine}
H.~Tran, C.~Kim, L.~Chen, A.~Chandrasekaran, R.~Batra, S.~Venkatram, D.~Kamal,
  J.~P. Lightstone, R.~Gurnani, P.~Shetty, M.~Ramprasad, J.~Laws, M.~Shelton
  and R.~Ramprasad, \emph{J. Appl. Phys.}, 2020, \textbf{128}, 171104\relax
\mciteBstWouldAddEndPuncttrue
\mciteSetBstMidEndSepPunct{\mcitedefaultmidpunct}
{\mcitedefaultendpunct}{\mcitedefaultseppunct}\relax
\EndOfBibitem
\bibitem[Chen \emph{et~al.}(2021)Chen, Pilania, Batra, Huan, Kim, Kuenneth, and
  Ramprasad]{chen2021polymer}
L.~Chen, G.~Pilania, R.~Batra, T.~D. Huan, C.~Kim, C.~Kuenneth and
  R.~Ramprasad, \emph{Mater. Sci. Eng. R Rep.}, 2021, \textbf{144},
  100595\relax
\mciteBstWouldAddEndPuncttrue
\mciteSetBstMidEndSepPunct{\mcitedefaultmidpunct}
{\mcitedefaultendpunct}{\mcitedefaultseppunct}\relax
\EndOfBibitem
\bibitem[Tran \emph{et~al.}(2024)Tran, Gurnani, Kim, Pilania, Kwon, Lively, and
  Ramprasad]{tran2024polymer}
H.~Tran, R.~Gurnani, C.~Kim, G.~Pilania, H.-K. Kwon, R.~Lively and
  R.~Ramprasad, \emph{Nat. Rev. Mater.}, 2024, \textbf{9}, 866–886\relax
\mciteBstWouldAddEndPuncttrue
\mciteSetBstMidEndSepPunct{\mcitedefaultmidpunct}
{\mcitedefaultendpunct}{\mcitedefaultseppunct}\relax
\EndOfBibitem
\bibitem[Gurnani \emph{et~al.}(2024)Gurnani, Shukla, Kamal, Wu, Hao, Kuenneth,
  Aklujkar, Khomane, Daniels, Deshmukh,\emph{et~al.}]{gurnani2024ai}
R.~Gurnani, S.~Shukla, D.~Kamal, C.~Wu, J.~Hao, C.~Kuenneth, P.~Aklujkar,
  A.~Khomane, R.~Daniels, A.~A. Deshmukh \emph{et~al.}, \emph{Nat. Commun.},
  2024, \textbf{15}, 6107\relax
\mciteBstWouldAddEndPuncttrue
\mciteSetBstMidEndSepPunct{\mcitedefaultmidpunct}
{\mcitedefaultendpunct}{\mcitedefaultseppunct}\relax
\EndOfBibitem
\bibitem[Cassola \emph{et~al.}(2022)Cassola, Duhovic, Schmidt, and
  May]{cassola2022machine}
S.~Cassola, M.~Duhovic, T.~Schmidt and D.~May, \emph{Compos. B Eng.}, 2022,
  \textbf{246}, 110208\relax
\mciteBstWouldAddEndPuncttrue
\mciteSetBstMidEndSepPunct{\mcitedefaultmidpunct}
{\mcitedefaultendpunct}{\mcitedefaultseppunct}\relax
\EndOfBibitem
\bibitem[Sharma \emph{et~al.}(2022)Sharma, Mukhopadhyay, Rangappa, Siengchin,
  and Kushvaha]{sharma2022advances}
A.~Sharma, T.~Mukhopadhyay, S.~M. Rangappa, S.~Siengchin and V.~Kushvaha,
  \emph{Arch. Comput. Methods Eng.}, 2022, \textbf{29}, 3341--3385\relax
\mciteBstWouldAddEndPuncttrue
\mciteSetBstMidEndSepPunct{\mcitedefaultmidpunct}
{\mcitedefaultendpunct}{\mcitedefaultseppunct}\relax
\EndOfBibitem
\bibitem[Matos \emph{et~al.}(2019)Matos, Pinho, and
  Tagarielli]{matos2019application}
M.~A. Matos, S.~T. Pinho and V.~L. Tagarielli, \emph{Carbon}, 2019,
  \textbf{146}, 265--275\relax
\mciteBstWouldAddEndPuncttrue
\mciteSetBstMidEndSepPunct{\mcitedefaultmidpunct}
{\mcitedefaultendpunct}{\mcitedefaultseppunct}\relax
\EndOfBibitem
\bibitem[Matos \emph{et~al.}(2019)Matos, Pinho, and
  Tagarielli]{matos2019predictions}
M.~Matos, S.~Pinho and V.~Tagarielli, \emph{Scr. Mater.}, 2019, \textbf{166},
  117--121\relax
\mciteBstWouldAddEndPuncttrue
\mciteSetBstMidEndSepPunct{\mcitedefaultmidpunct}
{\mcitedefaultendpunct}{\mcitedefaultseppunct}\relax
\EndOfBibitem
\bibitem[Altarazi \emph{et~al.}(2018)Altarazi, Ammouri, and
  Hijazi]{altarazi2018artificial}
S.~Altarazi, M.~Ammouri and A.~Hijazi, \emph{Comput. Mater. Sci.}, 2018,
  \textbf{153}, 1--9\relax
\mciteBstWouldAddEndPuncttrue
\mciteSetBstMidEndSepPunct{\mcitedefaultmidpunct}
{\mcitedefaultendpunct}{\mcitedefaultseppunct}\relax
\EndOfBibitem
\bibitem[Reddy \emph{et~al.}(2020)Reddy, Premasudha, Panigrahi, Cho, and
  Reddy]{reddy2020modeling}
B.~R.~S. Reddy, M.~Premasudha, B.~B. Panigrahi, K.-K. Cho and N.~G.~S. Reddy,
  \emph{Polym. Compos.}, 2020, \textbf{41}, 3208--3217\relax
\mciteBstWouldAddEndPuncttrue
\mciteSetBstMidEndSepPunct{\mcitedefaultmidpunct}
{\mcitedefaultendpunct}{\mcitedefaultseppunct}\relax
\EndOfBibitem
\bibitem[Zeng \emph{et~al.}(2019)Zeng, Hu, Zou, Zhang, and Sun]{zeng2019bp}
G.~S. Zeng, C.~Hu, S.~Zou, L.~Zhang and G.~Sun, \emph{Polym. Compos.}, 2019,
  \textbf{40}, 3923--3928\relax
\mciteBstWouldAddEndPuncttrue
\mciteSetBstMidEndSepPunct{\mcitedefaultmidpunct}
{\mcitedefaultendpunct}{\mcitedefaultseppunct}\relax
\EndOfBibitem
\bibitem[Kushvaha \emph{et~al.}(2020)Kushvaha, Kumar, Madhushri, and
  Sharma]{kushvaha2020artificial}
V.~Kushvaha, S.~A. Kumar, P.~Madhushri and A.~Sharma, \emph{J. Compos. Mater.},
  2020, \textbf{54}, 3099--3108\relax
\mciteBstWouldAddEndPuncttrue
\mciteSetBstMidEndSepPunct{\mcitedefaultmidpunct}
{\mcitedefaultendpunct}{\mcitedefaultseppunct}\relax
\EndOfBibitem
\bibitem[Ataeefard \emph{et~al.}(2019)Ataeefard, Mohammadi, and
  Saeb]{ataeefard2019intelligently}
M.~Ataeefard, Y.~Mohammadi and M.~R. Saeb, \emph{Polym. Sci. Ser. A}, 2019,
  \textbf{61}, 667--680\relax
\mciteBstWouldAddEndPuncttrue
\mciteSetBstMidEndSepPunct{\mcitedefaultmidpunct}
{\mcitedefaultendpunct}{\mcitedefaultseppunct}\relax
\EndOfBibitem
\bibitem[Weininger(1988)]{smiles}
D.~Weininger, \emph{J. Chem. Inf. Comput. Sci.}, 1988, \textbf{28},
  31--36\relax
\mciteBstWouldAddEndPuncttrue
\mciteSetBstMidEndSepPunct{\mcitedefaultmidpunct}
{\mcitedefaultendpunct}{\mcitedefaultseppunct}\relax
\EndOfBibitem
\bibitem[Rasmussen and Williams(2006)]{GPRBook}
\emph{Gaussian Processes for Machine Learning}, ed. C.~E. Rasmussen and
  C.~K.~I. Williams, The MIT Press, Cambridge, MA, 2006\relax
\mciteBstWouldAddEndPuncttrue
\mciteSetBstMidEndSepPunct{\mcitedefaultmidpunct}
{\mcitedefaultendpunct}{\mcitedefaultseppunct}\relax
\EndOfBibitem
\bibitem[Williams and Rasmussen(1995)]{GPR95}
C.~K.~I. Williams and C.~E. Rasmussen, \emph{Advances in Neural Information
  Processing Systems 8}, MIT Press, 1995\relax
\mciteBstWouldAddEndPuncttrue
\mciteSetBstMidEndSepPunct{\mcitedefaultmidpunct}
{\mcitedefaultendpunct}{\mcitedefaultseppunct}\relax
\EndOfBibitem
\bibitem[Glorot and Bengio(2010)]{glorot2010understanding}
X.~Glorot and Y.~Bengio, Proceedings of the thirteenth international conference
  on artificial intelligence and statistics, 2010, pp. 249--256\relax
\mciteBstWouldAddEndPuncttrue
\mciteSetBstMidEndSepPunct{\mcitedefaultmidpunct}
{\mcitedefaultendpunct}{\mcitedefaultseppunct}\relax
\EndOfBibitem
\bibitem[Ioffe and Szegedy(2015)]{Ioffe2015BatchShift}
S.~Ioffe and C.~Szegedy, \emph{32nd International Conference on Machine
  Learning, ICML 2015}, 2015, \textbf{1}, 448--456\relax
\mciteBstWouldAddEndPuncttrue
\mciteSetBstMidEndSepPunct{\mcitedefaultmidpunct}
{\mcitedefaultendpunct}{\mcitedefaultseppunct}\relax
\EndOfBibitem
\bibitem[Hinton(2012)]{hinton2012improving}
G.~Hinton, \emph{arXiv preprint arXiv:1207.0580}, 2012\relax
\mciteBstWouldAddEndPuncttrue
\mciteSetBstMidEndSepPunct{\mcitedefaultmidpunct}
{\mcitedefaultendpunct}{\mcitedefaultseppunct}\relax
\EndOfBibitem
\bibitem[Li \emph{et~al.}(2018)Li, Chen, Hu, and
  Yang]{Li2018UnderstandingShift}
X.~Li, S.~Chen, X.~Hu and J.~Yang, \emph{Proceedings of the IEEE Computer
  Society Conference on Computer Vision and Pattern Recognition}, 2018,
  \textbf{2019-June}, 2677--2685\relax
\mciteBstWouldAddEndPuncttrue
\mciteSetBstMidEndSepPunct{\mcitedefaultmidpunct}
{\mcitedefaultendpunct}{\mcitedefaultseppunct}\relax
\EndOfBibitem
\bibitem[Kingma and Ba(2014)]{Kingma2014Adam:Optimization}
D.~P. Kingma and J.~L. Ba, \emph{ICLR 2015}, 2014\relax
\mciteBstWouldAddEndPuncttrue
\mciteSetBstMidEndSepPunct{\mcitedefaultmidpunct}
{\mcitedefaultendpunct}{\mcitedefaultseppunct}\relax
\EndOfBibitem
\bibitem[{Matmerize, Inc.}()]{polymrize_url}
{Matmerize, Inc.}, \emph{PolymRize},
  \href{https://polymrize.matmerize.com/}{https://polymrize.matmerize.com/}\relax
\mciteBstWouldAddEndPuncttrue
\mciteSetBstMidEndSepPunct{\mcitedefaultmidpunct}
{\mcitedefaultendpunct}{\mcitedefaultseppunct}\relax
\EndOfBibitem
\bibitem[Huan \emph{et~al.}(2016)Huan, Boggs, Teyssedre, Laurent, Cakmak,
  Kumar, and Ramprasad]{Huan:review}
T.~D. Huan, S.~Boggs, G.~Teyssedre, C.~Laurent, M.~Cakmak, S.~Kumar and
  R.~Ramprasad, \emph{Prog. Mater. Sci.}, 2016, \textbf{83}, 236\relax
\mciteBstWouldAddEndPuncttrue
\mciteSetBstMidEndSepPunct{\mcitedefaultmidpunct}
{\mcitedefaultendpunct}{\mcitedefaultseppunct}\relax
\EndOfBibitem
\bibitem[Mannodi-Kanakkithodi \emph{et~al.}(2016)Mannodi-Kanakkithodi, Treich,
  Huan, Ma, Tefferi, Cao, Sotzing, and Ramprasad]{Arun:review}
A.~Mannodi-Kanakkithodi, G.~Treich, T.~D. Huan, R.~Ma, M.~Tefferi, Y.~Cao,
  G.~Sotzing and R.~Ramprasad, \emph{Adv. Mater.}, 2016, \textbf{28},
  6277--6291\relax
\mciteBstWouldAddEndPuncttrue
\mciteSetBstMidEndSepPunct{\mcitedefaultmidpunct}
{\mcitedefaultendpunct}{\mcitedefaultseppunct}\relax
\EndOfBibitem
\bibitem[Sharma \emph{et~al.}(2014)Sharma, Wang, Lorenzini, Ma, Zhu, Sinkovits,
  Pilania, Oganov, Kumar, Sotzing, Boggs, and Ramprasad]{Vinit_Nature}
V.~Sharma, C.~C. Wang, R.~G. Lorenzini, R.~Ma, Q.~Zhu, D.~W. Sinkovits,
  G.~Pilania, A.~R. Oganov, S.~Kumar, G.~A. Sotzing, S.~A. Boggs and
  R.~Ramprasad, \emph{Nat. Commun.}, 2014, \textbf{5}, 4845\relax
\mciteBstWouldAddEndPuncttrue
\mciteSetBstMidEndSepPunct{\mcitedefaultmidpunct}
{\mcitedefaultendpunct}{\mcitedefaultseppunct}\relax
\EndOfBibitem
\bibitem[Kuenneth \emph{et~al.}(2021)Kuenneth, Rajan, Tran, Chen, Kim, and
  Ramprasad]{kuenneth2021polymer}
C.~Kuenneth, A.~C. Rajan, H.~Tran, L.~Chen, C.~Kim and R.~Ramprasad,
  \emph{Patterns}, 2021, \textbf{2}, 100238\relax
\mciteBstWouldAddEndPuncttrue
\mciteSetBstMidEndSepPunct{\mcitedefaultmidpunct}
{\mcitedefaultendpunct}{\mcitedefaultseppunct}\relax
\EndOfBibitem
\bibitem[Gurnani \emph{et~al.}(2023)Gurnani, Kuenneth, Toland, and
  Ramprasad]{gurnani2023polymer}
R.~Gurnani, C.~Kuenneth, A.~Toland and R.~Ramprasad, \emph{Chem. Mater.}, 2023,
  \textbf{35}, 1560--1567\relax
\mciteBstWouldAddEndPuncttrue
\mciteSetBstMidEndSepPunct{\mcitedefaultmidpunct}
{\mcitedefaultendpunct}{\mcitedefaultseppunct}\relax
\EndOfBibitem
\end{mcitethebibliography}

\providecommand*{\mcitethebibliography}{\thebibliography}
\csname @ifundefined\endcsname{endmcitethebibliography}
{\let\endmcitethebibliography\endthebibliography}{}

\end{document}